\documentstyle[aps,prl,psfig,multicol]{revtex}

\begin{document}
\newcommand{\vecr}{{\bf r}}
\newcommand{\vecy}{{\bf y}}
\newcommand{\veck}{{\bf k}}
\newcommand{\vecv}{{\bf v}}
\newcommand{\vecE}{{\bf E}}
\newcommand{\vecj}{{\bf j}}
\newcommand{\kperp}{{{\bf k}_\perp}}
\newcommand{\sinch}{{\rm sinch}}
\newcommand{\cth}{{\rm cth}}
\newcommand{\bartau}{{{\bar \tau}_T}}
\renewcommand{\Re}{{\rm Re}}
\newcommand{\spc}{{\,\,\,\,\,\,\,\,}}
\newcommand{\bea}{\begin{eqnarray}}
\newcommand{\eea}{\end{eqnarray}}
\renewcommand{\[}{\begin{equation}}
\renewcommand{\]}{\end{equation}}
\newcommand{\bef}{\begin{figure}}
\newcommand{\ef}{\end{figure}}
\newcommand{\ie}{{\it i.e.}}
\newcommand{\eg}{{\it e.g.}}
\newcommand{\llabel}[1]{\label{#1}}
\newcommand{\eq}[1]{Eq.~(\ref{#1})}
\newcommand{\fig}[1]{Fig.~\ref{#1}}


\title{Universal distribution of transparencies in highly conductive
Nb/AlO$_x$/Nb junctions}

\author{Y. Naveh, Vijay Patel, D. V. Averin, K. K. Likharev, and J. E. Lukens}
\address{Department of Physics and Astronomy \\
State University of New York, Stony Brook, NY 11794-3800}
\date{\today}
\maketitle


\begin{abstract}
We report the observation of the universal distribution of transparencies,
predicted by Schep
and Bauer [Phys.\ Rev.\ Lett.\ {\bf 78}, 3015 (1997)] for dirty sharp
interfaces, in uniform
Nb/AlO$_x$/Nb junctions with high specific conductance ($10^8$
Ohm$^{-1}$cm$^{-2}$).
Experiments used the BCS density of states in superconducting niobium for
transparency distribution
probing. Experimental results for both the dc $I-V$ curves at
magnetic-field-suppressed supercurrent and the Josephson
critical current in zero magnetic field coincide remarkably well with
calculations based on the multimode theory of
multiple Andreev reflections and the Schep-Bauer distribution.
\end{abstract}

\begin{multicols}{2}

The basic characteristic of a mesoscopic conductor is
its set of transmission coefficients, or ``transparencies'', defined as
the eigenvalues of the transfer matrix connecting all incoming electronic
modes to
outgoing modes (for a thorough review, see Ref.~\cite{Beenakker 97}). The
set of
transparencies for a given conductor determines all its transport properties
including
dc current and broadband current noise.

In 1982 Dorokhov showed \cite{Dorokhov 82} that the distribution of
transparencies in diffusive conductors is
universal, \ie, does not depend on dimension, geometry, carrier
density, and other
sample-specific properties:
\[ \label{Dorokhov}
\rho (D) = {G \over 2 G_0} {1 \over D \sqrt{1 - D}},
\]
where $G$ is the average conductance and $G_0 = 2e^2 / h$. The universality
of \eq{Dorokhov} is
responsible in particular for the universal value $S_I(0) = 2eI/3$ of
shot noise in diffusive
conductors\cite{Beenakker 92,Sukhorukov 98} (here $I$ is the dc
current). This suppression of shot noise in comparison with its Schottky
value
has been observed experimentally\cite{Steinbach 96,Henny
97,Schoelkopf 97}
and may serve as an indirect confirmation of \eq{Dorokhov}.

Recently, Schep and Bauer showed \cite{Schep 97a+97b} that the distribution
of transparencies of a disordered interface is also universal,
but is given by an expression different from
\eq{Dorokhov} \cite{Ben}:
\[ \label{Schep}
\rho (D) = {G \over \pi G_0} {1 \over D^{3/2} \sqrt{1 - D}}.
\]
(This distribution leads to a shot noise value of $S_I(0) =
2eI / 2$\cite{Schep 97a+97b}).
In this work we report a strong experimental evidence that the transparency
distribution in
sub-nm-thick aluminum oxide barriers is very close to \eq{Schep} while being
substantially different from \eq{Dorokhov}.

The determination of the transparency distribution may be assisted by the
fact that due to
the BCS singularity in the density of states at the edges of the
superconductor energy gap
2$\Delta(T)$, both the Cooper-pair and quasiparticle transport through
Josephson junctions are highly nonlinear. In particular,
quasiparticle transfer strongly increases at the "gap voltage" $V_g = 2
\Delta(T)/e$, while below this
threshold the transport is dominated by multiple Andreev reflections (MAR)
\cite{Bratus 95,Averin
95,Bardas 97}, resulting in a pronounced sub-harmonic gap structure at $V_n
= V_g/n$. This structure is
very sensitive to the number and transparencies of the modes, and has been
successfully used
\cite{vanderPost 94,Scheer 97,Scheer 98} to determine transparencies of
atomic-size point contacts with a few propagating modes. However, as
will be shown below,
the fact that various features of MAR transport (the current jump at
$V = V_g$, the excess current at $V > V_g$, and subharmonic structure
at $V < V_g$) are sensitive to different ranges of the transparency
distribution
also allows probing of the distribution in junctions with a much
larger area (and
hence a very large number of propagating modes).

\begin{figure}
\narrowtext
\centerline{\hspace{-0.6cm} \psfig{figure=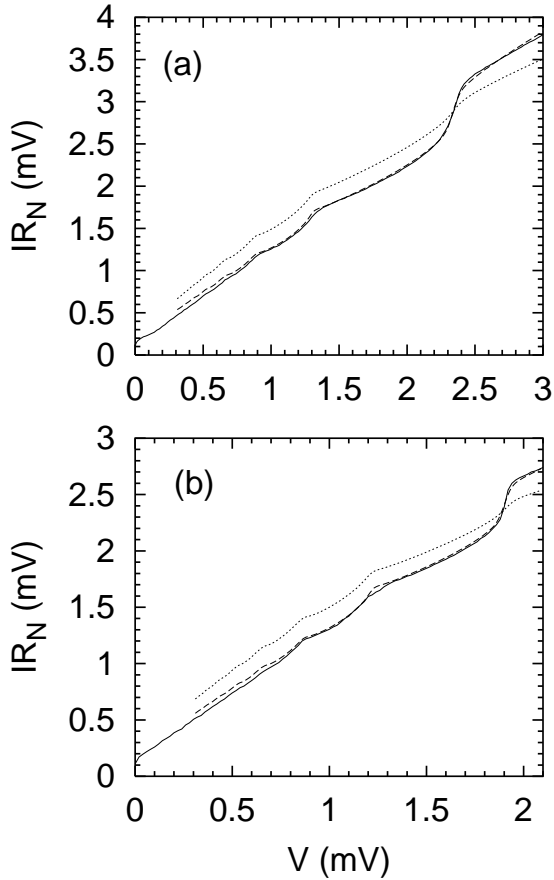,angle=-90,width=160mm}}
\caption{DC $I-V$ curves for two samples of area (a) $0.25 \mu$m$^2$ and
(b) $1.0 \mu$m$^2$. Solid lines: experimental results normalized with $1/G =
R_N =
R_N^{\rm expt}$. Dashed
lines: MAR theory using \eq{Schep}, (a) $R_N = 1.01 R_N^{\rm expt}$,
(b) $R_N = 1.04 R_N^{\rm expt}$. Dotted lines: MAR theory using
\eq{Dorokhov}, (a) $R_N =
0.82 R_N^{\rm expt}$, (b) $R_N = 0.84 R_N^{\rm expt}$.}
\label{1I-V}
\end{figure}

Our samples were made using in-situ-fabricated Nb/AlO$_x$/Nb trilayers which
were deposited on oxidized Si wafers
in a cryopumped vacuum system with a base pressure of $5\times 10^{-8}$ Torr. 
A dc magnetron sputtered 150-nm niobium base electrode was covered
by a 8-nm-thick aluminum film (also using dc sputtering). Without breaking
the vacuum, an AlO$_x$ layer was formed by
thermal oxidation with well controlled dry oxygen exposure. For the
junctions
discussed below (specific normal conductance
close to 10$^8$ Ohm$^{-1}$cm$^{-2}$), the oxidation was carried
out at 1.0 mTorr of O$_2$ for 10 minutes. After the trilayer
deposition was completed by sputtering of a 150-nm-thick Nb counter
electrode, junctions of various areas (from 0.25
to 1 $\mu$m$^2$) were formed by e-beam patterning  - for details,
see \cite{Bhushan 95,Patel 99}. The resulting junctions were highly uniform;
for example, for 1 $\mu$m$^2$ junctions the full spread of normal
conductance was below $\pm 3\%$; a slightly larger ($\pm 8\%$) spread of
deep-submicron
junctions may be readily explained by their area definition uncertainty.
The junction homogeneity was further confirmed by the fact that the
Josephson supercurrent could be almost
completely suppressed by a magnetic field corresponding to the insertion of
one flux quantum into the junction - see Fig. \ref{1I-V}.

\begin{figure}
\narrowtext
\centerline{\hspace{-0.6cm}
\psfig{figure=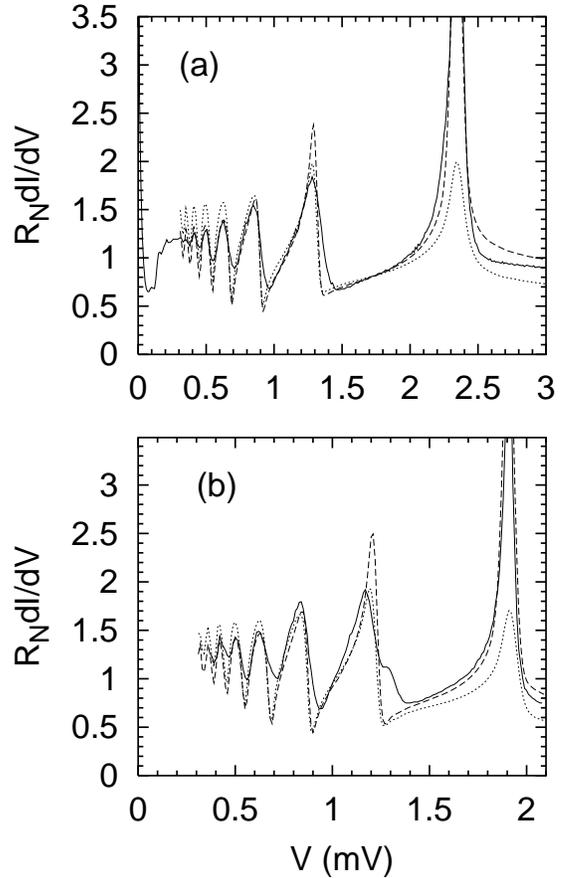,angle=-90,width=160mm}}
\caption{Differential conductance for the same samples, and with the same
fitting values of $R_N$, as in
\fig{1I-V}.}
\label{2dIdV-V}
\end{figure}

The solid lines in Fig. \ref{1I-V} show the experimental dc $I-V$ curves of
two
junctions of different area at a bath
temperature of $T_0 = 1.8$ K, with the Josephson critical current suppressed
by
a magnetic field parallel to the film plane, while the solid lines in Fig.
\ref{2dIdV-V} show the measured
differential conductances of the
same samples. These curves exhibit well pronounced sub-gap structure,
indicative of MAR transport. However, the conduction peak positions $V_n$
at higher voltages deviate gradually from the expected dependence
$1/n$, indicating self-heating of the samples. This heating was higher in
larger junctions and may be calculated with good accuracy using a simple
model where the main temperature gradient between the self-heated junction
and the bath is parallel to the substrate, in the niobium
electrodes adjacent to the junction; from
this picture one should expect overheating to scale roughly as the junction
area divided by its perimeter, in agreement with experiment.

\vspace{1.0cm}
\noindent

\begin{figure}
\narrowtext
\centerline{\hspace{-0.8cm}
\psfig{figure=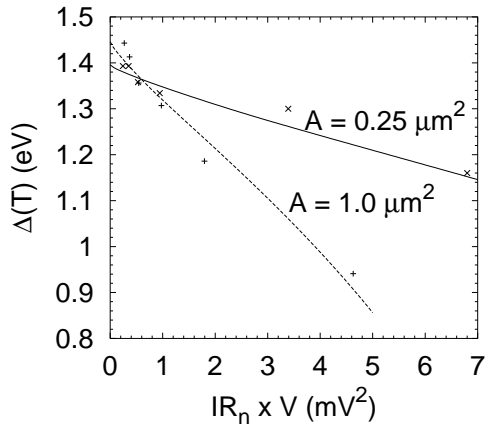,angle=-90,width=70mm}}
\caption{Dependence of the superconducting gap $\Delta(T)$ on the
dissipated power $P = IV$ for the same two samples as in Figs.
\ref{1I-V} and \ref{2dIdV-V}. Points: fit
to positions of the peaks in
\fig{2dIdV-V} (see the text). Lines: Calculated $\Delta(T)$ based on our
heating model and the BCS gap temperature dependence.}
\label{3Delta-IV}
\end{figure}

The points in \fig{3Delta-IV} show the dependence of the energy gap $\Delta
(T)$, where $T$ is the junction temperature
including self-heating, on the power $P = IV$ dissipated in the junction,
read off from the peak positions. Assuming the BCS temperature dependence of
the energy gap, the data show that at $V = V_g(T)$ the junction temperature 
is close to 6.2 K and 7.6 K for, respectively, the $0.25 \mu$m$^2$ and
$1.0 \mu$m$^2$ junctions.  The solid and dashed lines in
\fig{3Delta-IV} show the theoretical dependence of the gap on $P$
assuming a specific heating
model. In this model we approximate the experimental temperature dependence
of the heat
conductance $\kappa(T)$ of superconducting niobium \cite{Kes 74} with a
one-parameter parabolic expression.
The resulting dependences provide a reasonably
good, smooth interpolation of the experimental heating data.

\vspace{1.0cm}

Dashed curves in Figures \ref{1I-V} and \ref{2dIdV-V} show the 
theoretical dependences which were obtained by averaging the 
results of the MAR theory \cite{Averin 95,Bardas 97} over the
distribution of transparencies given by
\eq{Schep} \cite{Brinkman 99}. Apart from the incorporation of the 
heating model described above, two additional minor adjustments
were made when calculating these curves. First, the assumed value 
of normal resistance $R_N$ was allowed to differ
by a few percent (within experimental uncertainty \cite{Rn}) 
from the experimentally
determined value $R_{N}^{\rm expt}$.
Second, we have introduced a Gaussian distribution of
$\Delta(T)$ with an r.m.s.\ spread of 2\%. Such a
spread is typical for any superconductor junction and may be readily
attributed to the anisotropy of the
superconducting gap in polycrystalline electrode films.

We believe that the agreement of the experimental data and theoretical
curves based on the Schep-Bauer
distribution is
remarkable. In order to see that this agreement could not result from the
fitting procedure described above, dotted
curves in Figs. \ref{1I-V} and \ref{2dIdV-V}
show the results of our best attempt to fit the data with the MAR theory
results averaged using the Dorokhov distribution (with a
similar account of self-heating). It is
evident that these curves are rather far from the data. Moreover, in this
best
fitting attempt
we have selected the values of $R_N$ rather distant from $R_{N}^{\rm expt}$
(see
the Fig.~\ref{1I-V} caption); if the latter
values were used, the theoretical lines
would pass considerably higher than the experimental plots. This is
immediately visible from the values of the ``excess current''
defined as $I_{\rm exc} = I(V) - G V$ at $V \gg V_g$.  Averaging the results
of Ref. \cite{Averin 95} for $T = 0$ with the Schep-Bauer distribution
\eq{Schep}, we get
\[
I_{\rm exc} = {G \Delta(0) \over e} \pi \left( {7 \over 4} - \sqrt{2}
\right) \approx 1.055 {G \Delta(0) \over e}.
\]
while for the Dorokhov distribution the excess current is substantially
higher, $I_{\rm exc} \approx 1.467 {G \Delta(0)/e}$  \cite{Artemenko}.

\begin{figure}
\narrowtext
\centerline{\hspace{-0.5cm} \psfig{figure=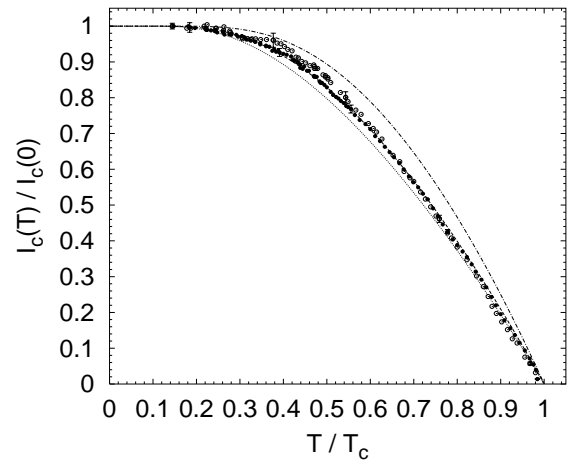,angle=-90,width=80mm}}
\caption{Temperature dependence of the critical current $I_c$ in the
two samples of Figs.~\ref{1I-V}--\ref{3Delta-IV}. Open circles: $A = 0.25
\,\,
\mu^2$; filled circles: $A = 1.0 \,\, \mu^2$; dashed-dotted line:
the Ambegaokar--Baratoff result for tunnel junctions; dashed line: MAR
theory
using \eq{Schep}; dotted line: MAR theory using \eq{Dorokhov}.}
\label{4Jc-T}
\end{figure}

Another (though less spectacular) evidence of the validity of the
MAR theory combined with the Schep-Bauer distribution
comes from the temperature dependence of the Josephson critical current (in
zero magnetic field), which is also sensitive
to the mode transparency distribution. Points in Fig.~\ref{4Jc-T} show the
measured temperature dependence of the critical current
in the same samples as in Figs.~\ref{1I-V}--\ref{3Delta-IV}.
Lines show the Ambegaokar-Baratoff dependence for tunnel junctions 
(upper curve), the MAR theory using the Dorokhov distribution 
(lower curve), and the MAR theory using the
Schep-Bauer distribution (middle curve) \cite{ave}. It is evident
that the experimental data agree with the Schep-Bauer distribution.

The absolute value of the low-temperature critical current is different in
all
three models used in Fig.~\ref{4Jc-T}: $I_c(0)/(G\Delta(0)/e)=1.57\, ;
1.92\, ; 2.08$ for, respectively, a tunnel junction, disordered barrier
with the Schep-Bauer distribution of transparencies, and SNS junctions
with Dorokhov's distribution \cite{ko1}. The observed value lies
within a few percent of the prediction for the Schep-Bauer distribution;
the difference can again be attributed to the error in the normal resistance
definition \cite{Rn}.

The fact that transport in disordered AlO$_x$ barriers of finite thickness
(of the order of 1 nm, \ie, much thicker than the Fermi
wavelength $\lambda_F$ in the junction electrodes) is described by
\eq{Schep}
so well may seem rather surprising, since its derivation in Ref. \cite{Schep
97a+97b}
assumed that the barrier is a
strongly-disordered region with thickness $d$ much smaller than $\lambda_F$.
However, this
distribution may be derived from a different model which does not rely
on this assumption.

It is well known that
resonant tunneling through a single localized site leads to the following
transparency:
\[ \label{rt}
D = {1 \over [(\epsilon - \epsilon_F)/\Gamma]^2+ \cosh^2 (x/a) } ,
\]
where $\epsilon$ is the state energy, $\epsilon_F$ is the Fermi level
in the junction electrodes, $\Gamma$ is the tunneling width for a site in
the barrier center, $x$ is the site deviation from the center, and $a$ is
the localization
radius. If $\Gamma$ is so large that the first term in the denominator is
unimportant, for a system with a uniform spatial distribution of sites,
\eq{rt} immediately gives the Dorokhov distribution. On the other hand, if
$x \approx 0$ while the spread of $\epsilon$
is much broader than $\Gamma$, \eq{rt} yields the distribution (\ref{Schep}).
This condition applies to strongly disordered barriers like ours
\cite{AlO}
if their thickness $d$ is smaller than $a$ (though possibly much larger 
than $\lambda_F$) and their spread of atomic localization energies is 
larger than $\Gamma$.

In summary, the excellent agreement between the experimental data
and the results of the MAR theory combined with the Schep-Bauer
distribution of transparencies provides very strong evidence that our
Nb/AlO$_x$/Nb junctions are well described by this distribution, while
being far from, e.g., the Dorokhov distribution. Since 
the distribution (\ref{Schep}) has the same universal nature as
\eq{Dorokhov}, this result is of considerable general importance
for mesoscopic physics \cite{previous}.

Our result is also of substantial importance for applications. It shows
that transport in niobium-trilayer Josephson junctions
with high specific conductance and hence high critical current density
(up to at least 200 kA/cm$^2$)
may be due to a fundamental mechanism rather than rare defects such as
pinholes, etc. This gives every hope that these
junctions may be even more reproducible than in our first experiments:
assuming that the barrier transparency
is only correlated at distances of the order of its thickness (about 1 nm),
we may estimate that the minimum
r.m.s. spread of
the critical current is below 1\% even for deep-submicron junctions.
Together with the fact that such junctions are intrinsically overshunted
\cite{Patel 99}, this makes them
uniquely suitable for several important applications in superconductor
electronics, including ultrafast
digital RSFQ circuits of very high integration scale - see, e.g., Ref.
\cite{Likh 00}.

We would like to thank A. A. Golubov and Yu. V. Nazarov for useful
discussions. This work was supported in part by DoD and NASA via JPL, and by
ONR.

\end{multicols}
\end{document}